\documentclass[twocolumn,amssymb,fleqn]{revtex4} 
\usepackage{epsfig,amssymb,amsmath,graphicx,subfigure,hyperref  }  
\usepackage{color}

\usepackage{multirow}
\usepackage{tabularx}
\usepackage{dsfont}

\begin{document}

\title{Command of Collective Dynamics by Topological Defects in Spherical Crystals}
\author{Zhenwei Yao}
\email{zyao@sjtu.edu.cn}
\affiliation{School of Physics and Astronomy, and Institute of Natural Sciences,
Shanghai Jiao Tong University, Shanghai 200240, China}
\begin{abstract} 
Directing individual motions of many constituents to coherent dynamical state is
  a fundamental challenge in multiple fields. Here, based on the spherical
  crystal model, we show that topological defects in particle arrays can be a
  crucial element in regulating collective dynamics. Specifically, we
  highlight the defect-driven synchronized breathing modes around disclinations
  and collective oscillations with strong connection to disruption of
  crystalline order. This work opens the promising possibility of an
  organizational principle based on topological defects, and may inspire new
  strategies for harnessing intriguing collective dynamics in extensive
  nonequilibrium systems. 
  \end{abstract}

\maketitle

Understanding self-organization of individual motions of many constituents into
coherent motions is a fundamental and practical problem in multiple fields
ranging from many-body problems~\cite{calogero2003classical, bruus2004many},
active
matter~\cite{vicsek1995novel,marchetti2013hydrodynamics,keber2014topology,
nguyen2014emergent} to biological processes~\cite{bray2000cell, Wang2011}.
Highly ordered collective motions in far-from-equilibrium many-body systems can
be driven by active
processes~\cite{marchetti2013hydrodynamics,keber2014topology}, physical
interactions~\cite{chattopadhyay2009effect,
nitzan2017revealing,mandal2018magnetic}, and exchange of
signals~\cite{bray2000cell,helbing2000simulating,chen2017weak}. Recently,
carefully designed lattice structures have been used to guide collective motions
of particles~\cite{yao2017emergent}, and host exotic unidirectional sound
modes~\cite{yang2015topological, shankar2017topological}. These studies suggest
an organizational principle based on establishing lattice structure among
constituents. While the connection of particle density and collective motion has
been extensively studied in various active matter
systems~\cite{zhang2010collective,vicsek2012collective,sumino2012large,
schaller2013topological}, the question of how the crystallographic structures in
particle arrays, particularly the defects therein, regulate the collective
dynamics has not yet been fully explored. Since collective motions can be
characterized by singularities in the velocity vector
field~\cite{schaller2013topological}, elucidating the interplay of the 
of defects in particle arrays and the vector field may inspire new strategies
for harnessing intriguing collective dynamics in extensive nonequilibrium
systems.

The spherical crystal model provides a suitable tool to address these
questions.  A spherical crystal refers to a two-dimensional crystal lattice
wrapping the entire sphere. Crystallographic defects, known as topological
defects, are inevitable in the spherical crystal as demanded by topological
constraints~\cite{bowick2002crystalline, Bausch2003e, nelson2002defects}.
Repulsive point particles on the sphere spontaneously form a crystal.
Determining its ground state is a 100-year-old puzzle known as the Thomson
problem~\cite{Thomson1904}, which has strong connections with virus
morphology~\cite{caspar1962physical,lidmar2003virus} and various geometric frustrations of
condensed matters~\cite{dinsmore2002colloidosomes,
Bausch2003e,yao2017topological, irvine2010pleats, azadi2014emergent, Mehta2016}.  In our
model, we introduce dynamics by imposing random disturbance to the lowest-energy
spherical crystal composed of long-range repulsive particles. By numerically
integrating the equations of motion at high precision, we reveal highly
symmetric singularity structures in the velocity vector field. The crucial
element for shaping such a well-organized velocity field is the synchronized
breathing modes around the twelve disclinations. We further identify a
collective oscillation mode that is closely related with disruption of
crystalline order. These two kinds of collective dynamical modes are generic in
spherical crystals of distinct symmetries at varying strength of disturbance.
These results demonstrate how dynamical order can be achieved by topological
defects in particle arrays, and may have implications in commanding the
nonequilibrium dynamics of active matters.

In our model, we consider a collection of point particles confined on the sphere
interacting by the Coulomb potential $V(r) = \beta/r$, where $r$ is the
Euclidean distance between two particles, and $\beta$ is a constant. According
to the Euler's theorem, topological defects are inevitable in the spherical
crystal~\cite{nelson2002defects}. The elementary topological defects in
two-dimensional triangular lattices are $n$-fold disclinations, which are
vertices whose coordination number $n$ is deviated from 6. Euler's theorem
states that on a closed triangulated surface $M$, 
\begin{eqnarray}
  \sum_{i} q_i = 2\pi\chi(M),
  \label{Euler}
  \end{eqnarray}
where $q_i$ is the topological charge of the vertex
$i$, and $\chi(M)$ is Euler's characteristic~\cite{nelson2002defects}.
$q_i=(6-n)\pi/3$ for a vertex with coordination number $n$.
$\chi(M)=2$ for the sphere, so topological defects are inevitable in the
spherical crystal. Following the Caspar-Klug scheme, we construct the basic structure of
the spherical crystal whose twelve fivefold disclinations are located at the
vertices of an inscribed icosahedron~\cite{caspar1962physical}. All possible
crystal lattices in the Caspar-Klug construction are represented by a pair of
non-negative integers $(p, q)$. The $q$ value reflects distinct symmetries of
the spherical crystal.

We introduce dynamics by imposing random disturbance to the lowest-energy
configuration
of the spherical crystal obtained by the steepest descent
method~\cite{yao2013topological,yao2016electrostatics}. Specifically, we pull
each particle away from its balance position by a displacement $\delta
\vec{x}$.
$\delta \vec{x} = \Gamma a
(\cos\alpha, \sin\alpha)$ in the unit basis vectors $\{\hat{e}_{\theta},
\hat{e}_{\phi} \}$. $a$ is the lattice spacing. $\Gamma a$ and $\alpha$ are the magnitude and direction of
the particle displacement. $\alpha$ is a uniform random variable in $[0, 2\pi)$.
The ensuing evolution of the system is governed by the following equations of
motion in spherical coordinates:  
\begin{eqnarray}
&&mR^2 \ddot{\theta}_i = mR^2 \dot{\phi}_i^2 \sin\theta_i \cos\theta_i - 
  \sum_{j \neq i}\frac{\partial V(r_{ij})}{\partial \theta_i},\nonumber \\
&&mR^2 \frac{d}{dt}\left( \sin^2\theta_i \dot{\phi}_i \right) = - \sum_{j \neq
  i}\frac{\partial V(r_{ij})}{\partial \phi_i}. \label{eom}
\end{eqnarray}
We numerically integrate Eqs.(\ref{eom}) for the particle trajectories with
various given initial conditions at high precision (see Supplemental Material for technical
details). The approach based on the equations of motion allows us to explore the
regime of large disturbance that is beyond the scope of perturbation analysis.
The length, mass and time are measured in the units of the spherical radius $R$,
particle mass $m$, and $\tau_0 = R\sqrt{m/\epsilon_0}$, where $\epsilon_0 =
\beta/R$.

\begin{figure}[t]  
\centering 
\subfigure[]{
\includegraphics[width=1.56in]{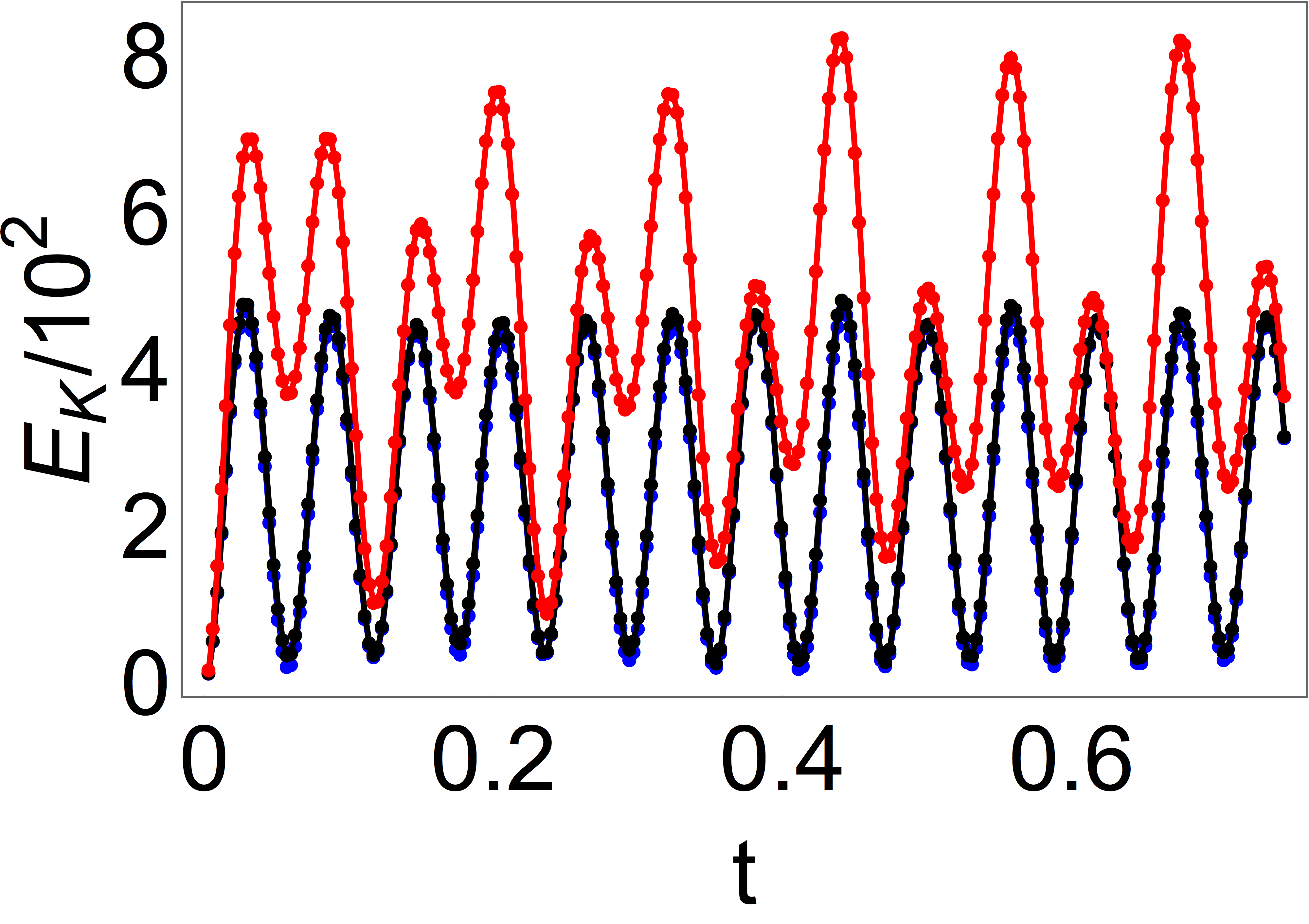}}  
\hspace{-0.02in} 
\subfigure[]{
\includegraphics[width=1.7in]{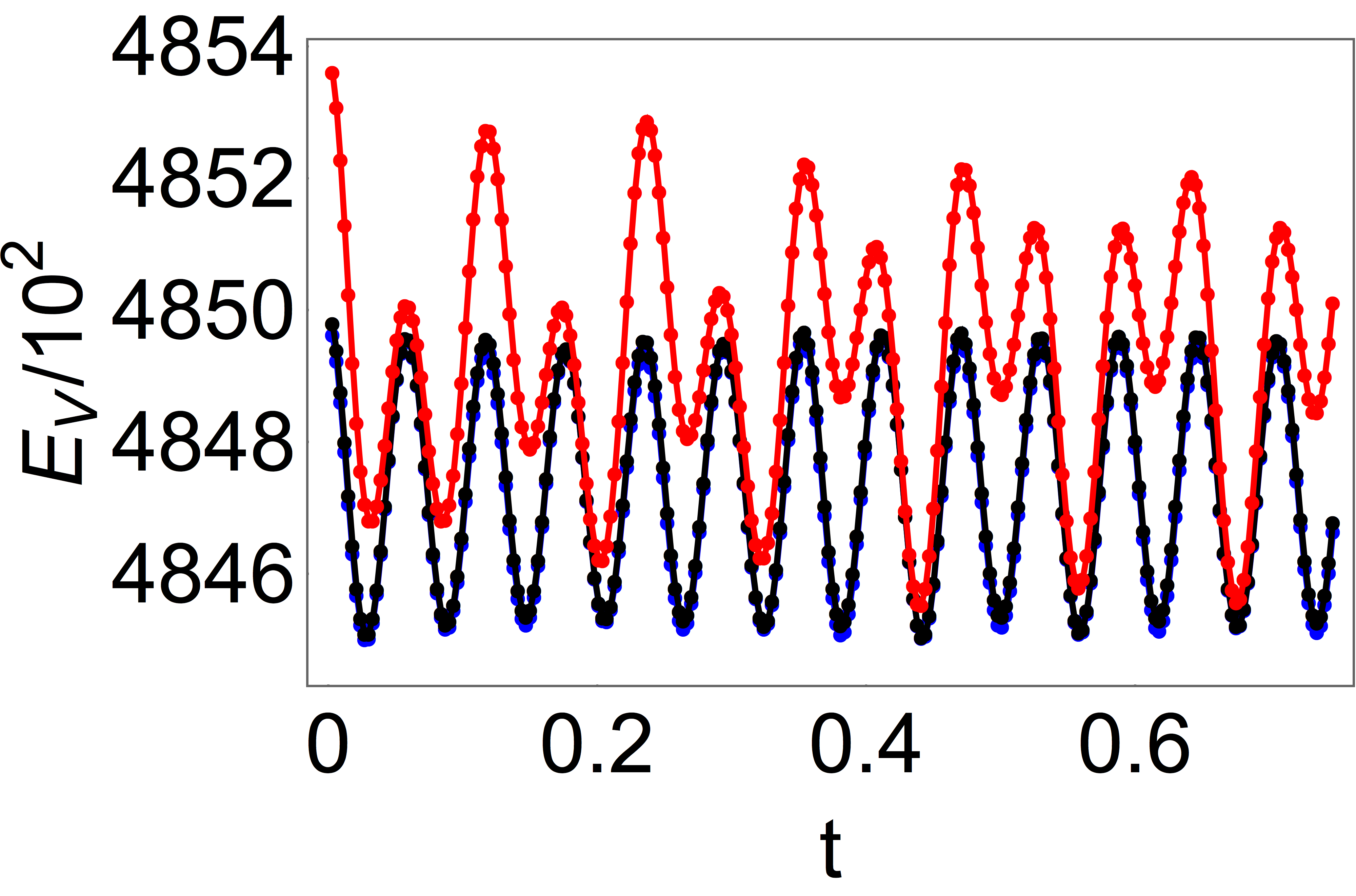}}  
\caption{Evolution of total kinetic and potential energies of disturbed
  spherical crystals. The total energy is
  well conserved. The amplitude of disturbance $\Gamma=1\%$
  (blue), $10\%$ (black), and $40\%$ (red). $(p, q) = (10, 0)$. 
    }
\label{energy}
\end{figure}

We first track the temporally varying total kinetic and potential energies, as
shown in Fig.~\ref{energy}. In the spherical system, the relation between the
kinetic and potential energies does not conform to the Virial theorem because of
the extra contribution from the constraint of the sphere to the force on a
particle~\cite{goldstein2011classical}. From Fig.~\ref{energy}, we see that for
small $\Gamma$, where the initial displacements of the particles from their
balance positions are small in comparison with the lattice spacing, the energy
curves are sinusoidal. With the increase of $\Gamma$, the energy curves exhibit
double-peak structures as shown in the red curves for $\Gamma=40\%$.  Note that
in even longer simulation time (up to 1.5 million simulation steps; $t=1.5$) the
energy curves still exhibit oscillating behaviors featured with the double-peak
structure, and the crystalline order of the system is well preserved. The
periodic oscillation of the energy curves suggests ordered dynamical modes
underlying the particle motions.

Prior to examining the dynamical modes on the sphere, we first present general
discussions about the topology of a vector field on the sphere.  As a classical problem of
differential topology, an even-dimensional sphere admits no regular tangent
vector field, and singularities are inevitable as a topological
constraint~\cite{aminov2000geometry}. A singularity at point $p$ in the vector
field $V$ is
characterized by its index $\textrm{Ind}_p V$: 
\begin{eqnarray}
  \textrm{Ind}_p V = \frac{1}{2\pi} \oint_{\gamma} d\theta,
\end{eqnarray}
where $\theta$ specifies the
direction of the vector field along a closed contour $\gamma$. $\textrm{Ind}_p V \in
\mathds{Z}$. Remarkably, there is a deep connection between vector fields and topology
illustrated by the Poincar$\acute{e}$-Hopf index
theorem~\cite{renteln2013manifolds}. It states that, over a compact manifold $M$
without boundary, regardless of the chosen vector field $V$,
\begin{eqnarray}
  \sum_{p\in M} \textrm{Ind}_p V = \chi(M).
  \label{PH}
\end{eqnarray}  
Applying Eq.(\ref{PH}) on the 2-sphere, we obtain the hairy ball theorem: $\sum_{p\in
\mathds{S}^2} \textrm{Ind}_p V = 2$. The name of this theorem is related to the
fact that it is impossible to comb the hairs without creating a cowlick.
Equation~(\ref{PH}) dictates a topological constraint on the configuration of
the vector
field on the sphere. It is natural to ask how the defects in the spherical
crystal will shape the velocity field and regulate the topologically required
singularities therein.

\begin{figure}[t]  
\centering 
\includegraphics[width=3in]{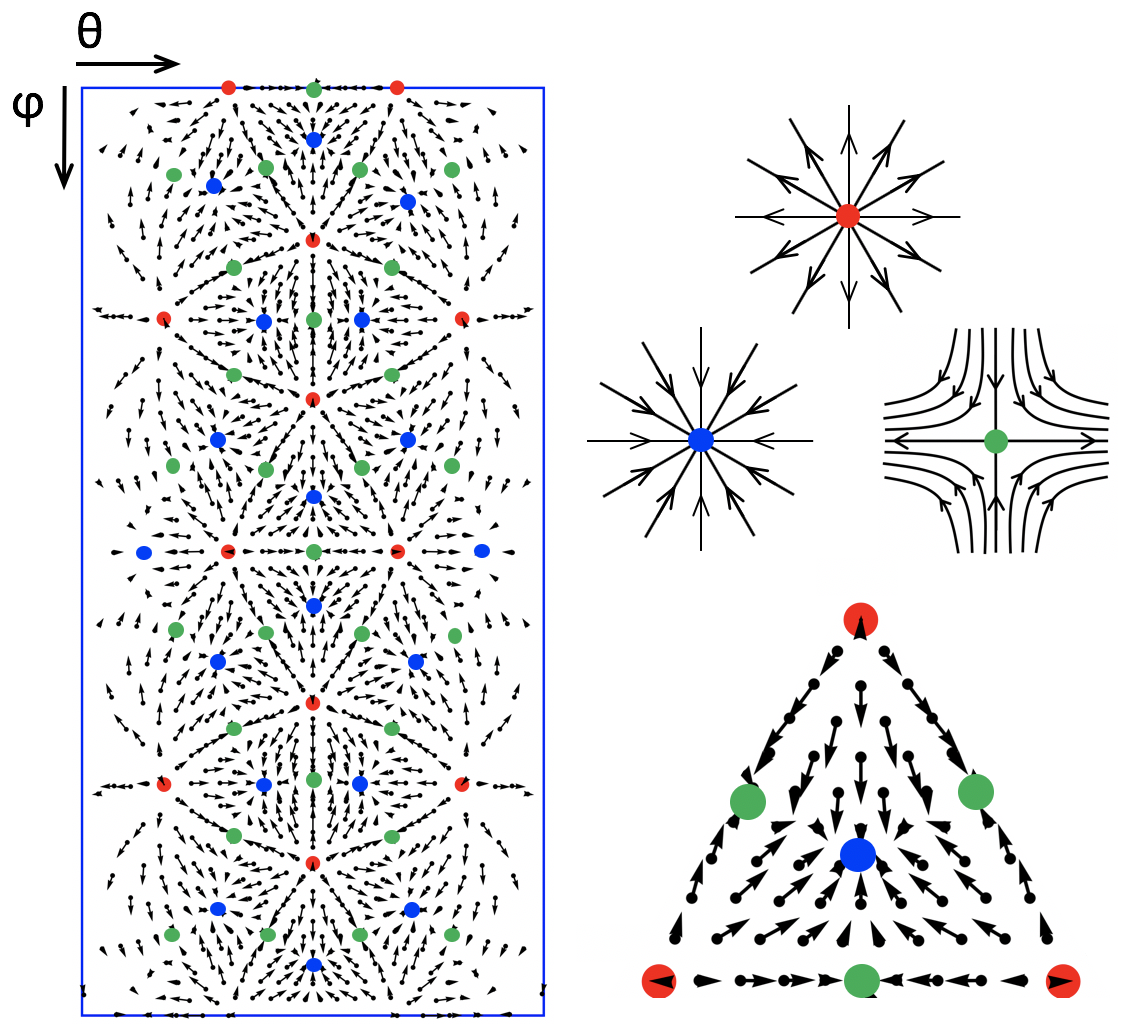}
  \caption{Highly symmetric velocity vector field regulated by the twelve
  disclinations (indicated by red dots) in the spherical crystal. The particle
  configuration is represented in the spherical coordinates $(\theta, \phi)$.
  The entire velocity field is equally compartmented into identical subfields
  in the spherical triangles spanned by three neighboring disclinations; a
  zoomed-in plot of a subfield is shown in the lower right panel. The three types
  of singularities are also schematically shown.  $\Gamma=1\%$.  $t=0.735$.
  $(p, q) = (10, 0)$.    }
\label{v_field}
\end{figure}

In Fig.~\ref{v_field}, we present a typical snapshot of the instantaneous velocity
vector field. The red dots are the preexistent fivefold disclinations in the
spherical crystal; they are located at the vertices of an inscribed icosahedron.
Remarkably, the entire velocity field is equally compartmented into identical subfields in
the spherical triangles spanned by three neighboring disclinations. We identify
three types of singularities in the velocity field: sources (red dots),  sinks
(blue dots), and saddle points (green dots), whose indices are $+1$, $+1$, and
$-1$, respectively. The entire velocity field is characterized by the highly
symmetric arrangement of these singularities. Specifically, the 20 blue dots and
30 green dots are located at the vertices of the inscribed dodecahedron and
icosidodecahedron, respectively. We also present the evolution of the velocity
field and the slight displacement of the singularity structure in Supplemental
Material. We count the total index value to be exactly $+2$. Notably, the vector
field configuration in Fig.~\ref{v_field} has been used to prove the hairy ball
theorem~\cite{aminov2000geometry}. For spherical crystals with nonzero $q$, the
velocity fields exhibit identical defect structure as in Fig.~\ref{v_field}; the
typical velocity fields for $q=2$, 3, and 6 are presented in Supplemental
Material. It indicates that the highly symmetric singularity structure is a
generic feature in the velocity field of the spherical crystals.

\begin{figure}[t]  
\centering 
\subfigure[]{
\includegraphics[width=1.62in]{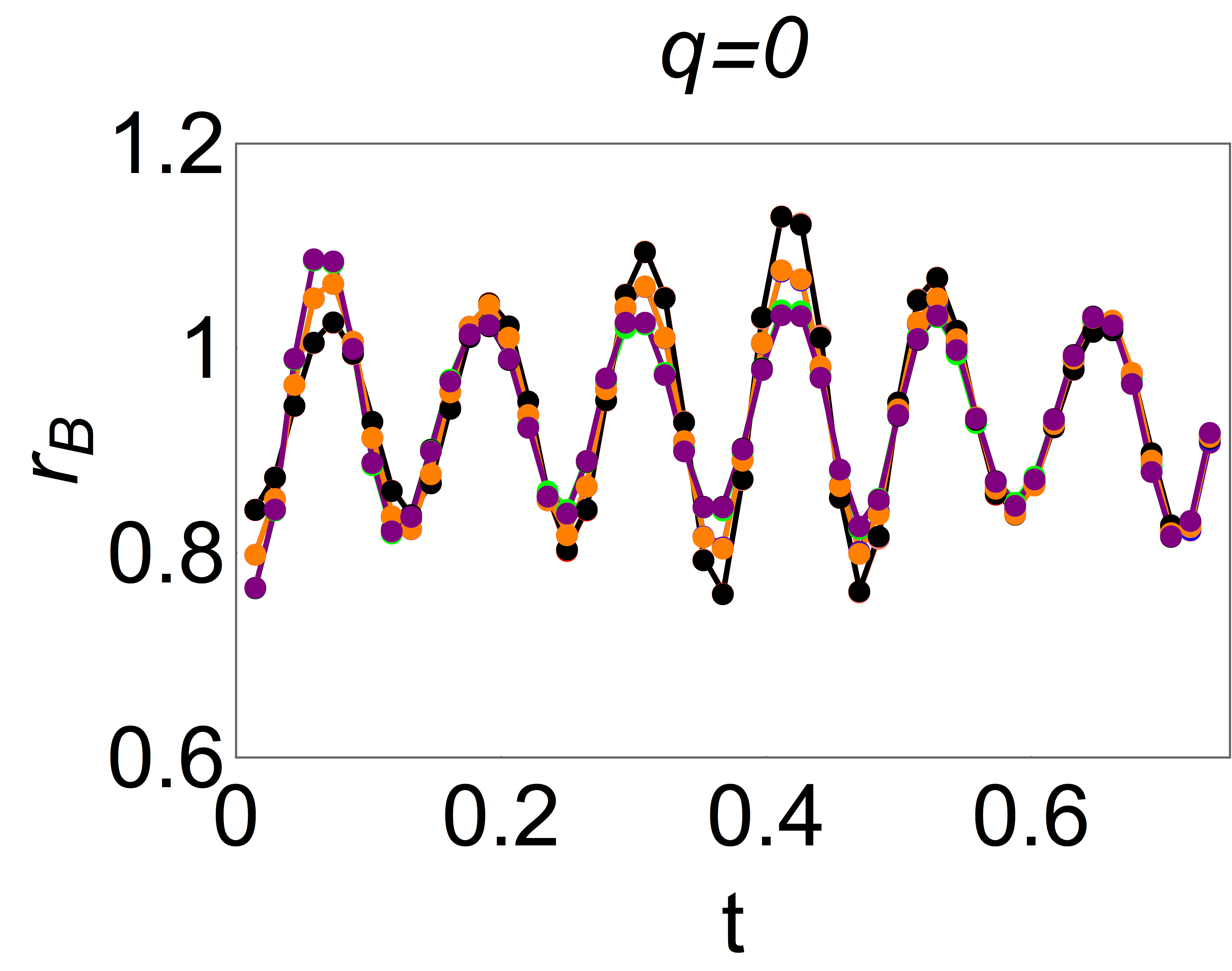}}  
\hspace{-0.02in} 
\subfigure[]{
\includegraphics[width=1.62in]{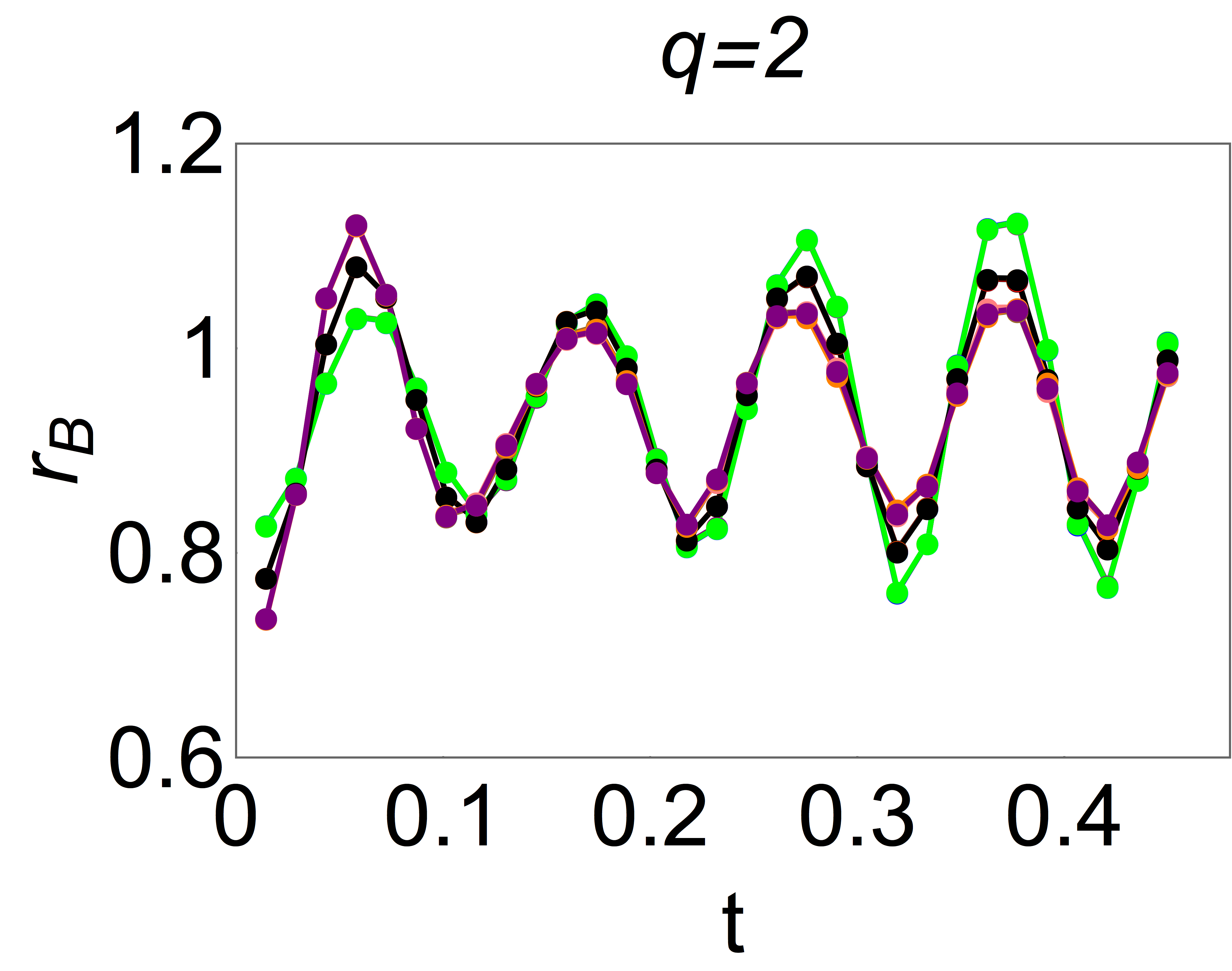}}  
\subfigure[]{
\includegraphics[width=1.62in]{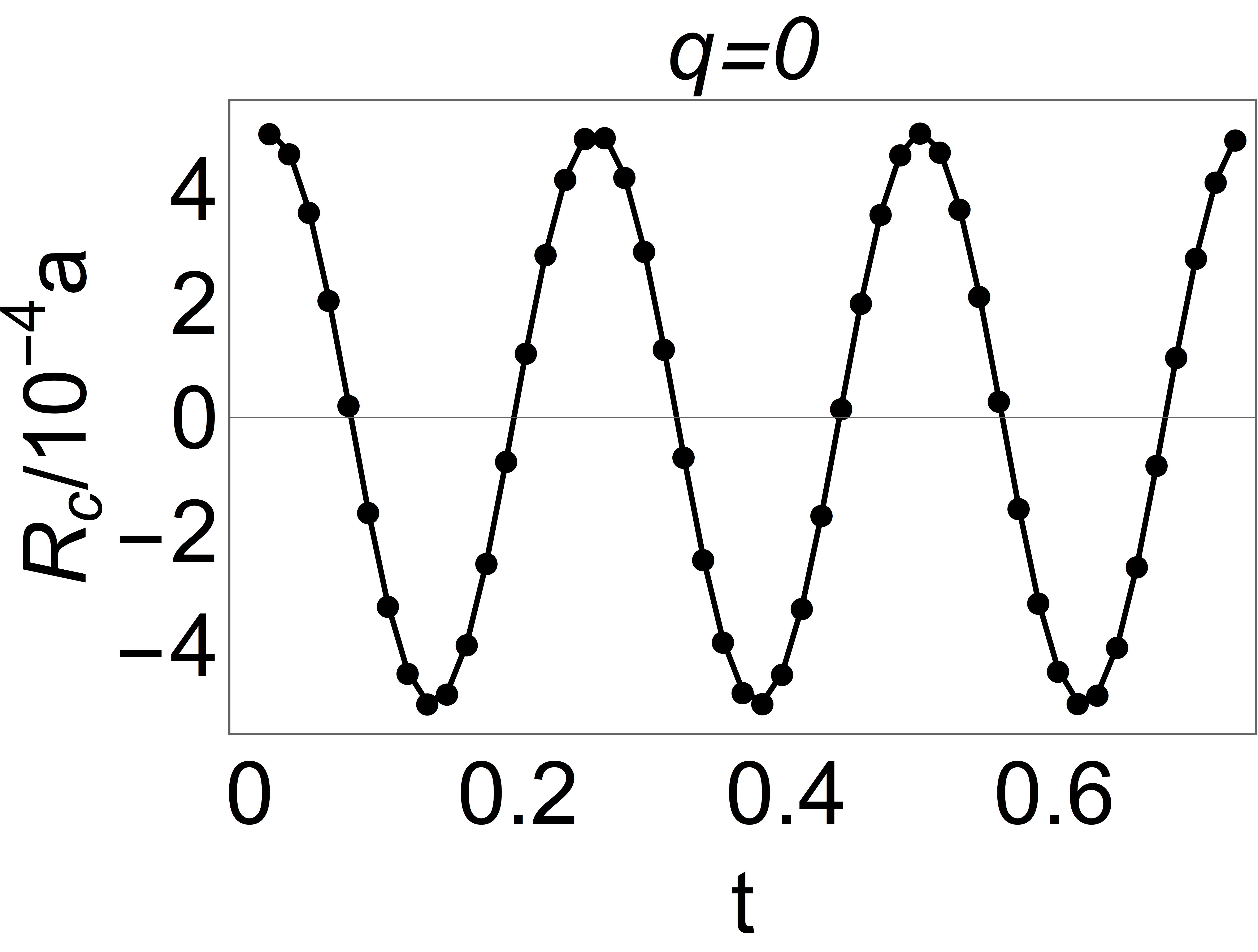}}  
\hspace{-0.02in} 
\subfigure[]{
\includegraphics[width=1.62in]{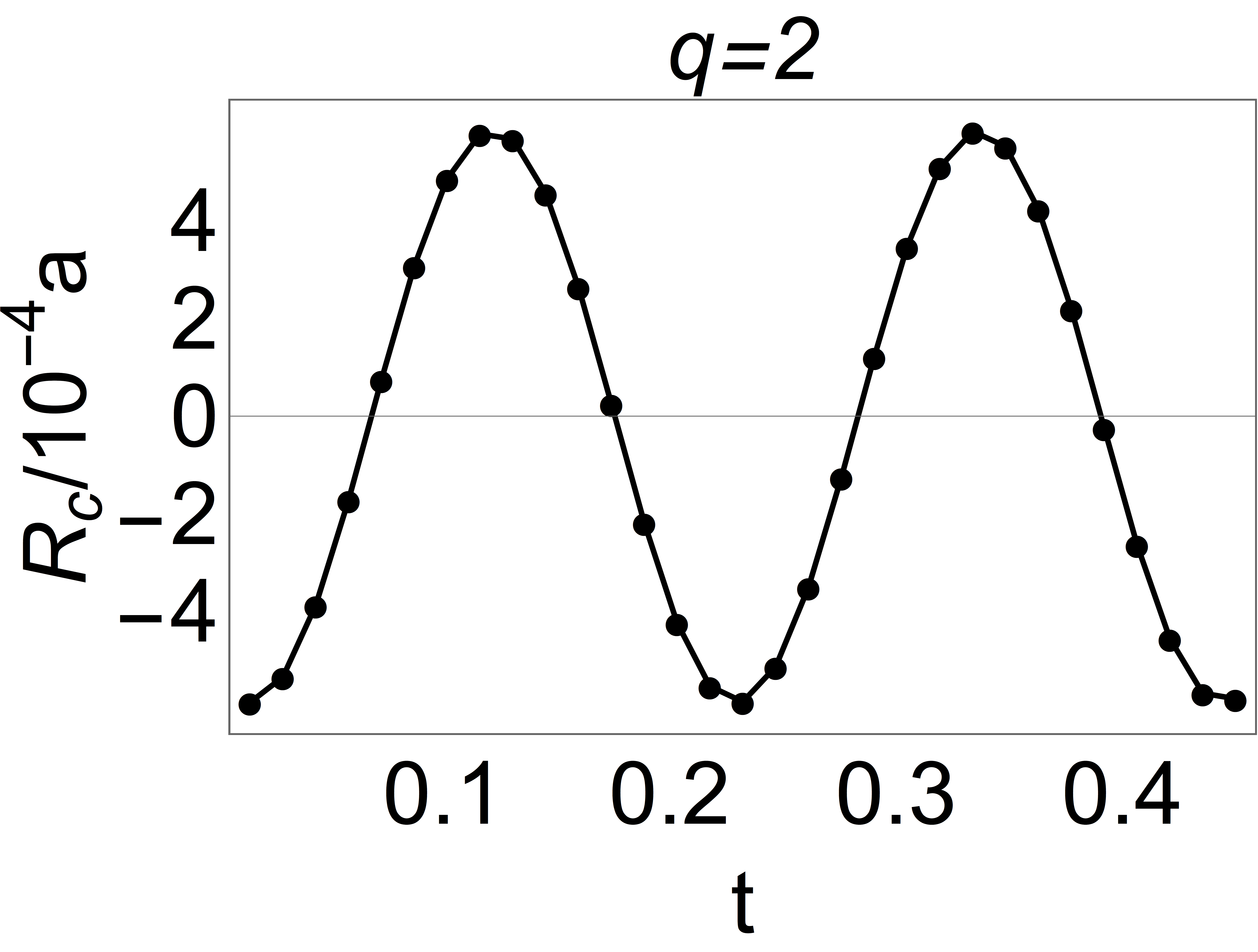}}  
\caption{Two kinds of dynamical modes revealed in the collective dynamics of the
  spherical crystals with distinct symmetries.  (a) and (b) show the synchronization of the breathing modes
  around the twelve disclinations.  $r_B$ is the average distance from the
  disclination to the five neighboring particles. (c) and (d) show the
  collective oscillation mode.  $R_c$ is the location of the center of mass on
  the z axis. The amplitude of oscillation is only at the order of
  ten-thousandth of the lattice spacing $a$. $\Gamma=1\%$.  $p=10$.
    }
\label{modes}
\end{figure}

\begin{figure*}[th]  
\centering 
\includegraphics[width=6in]{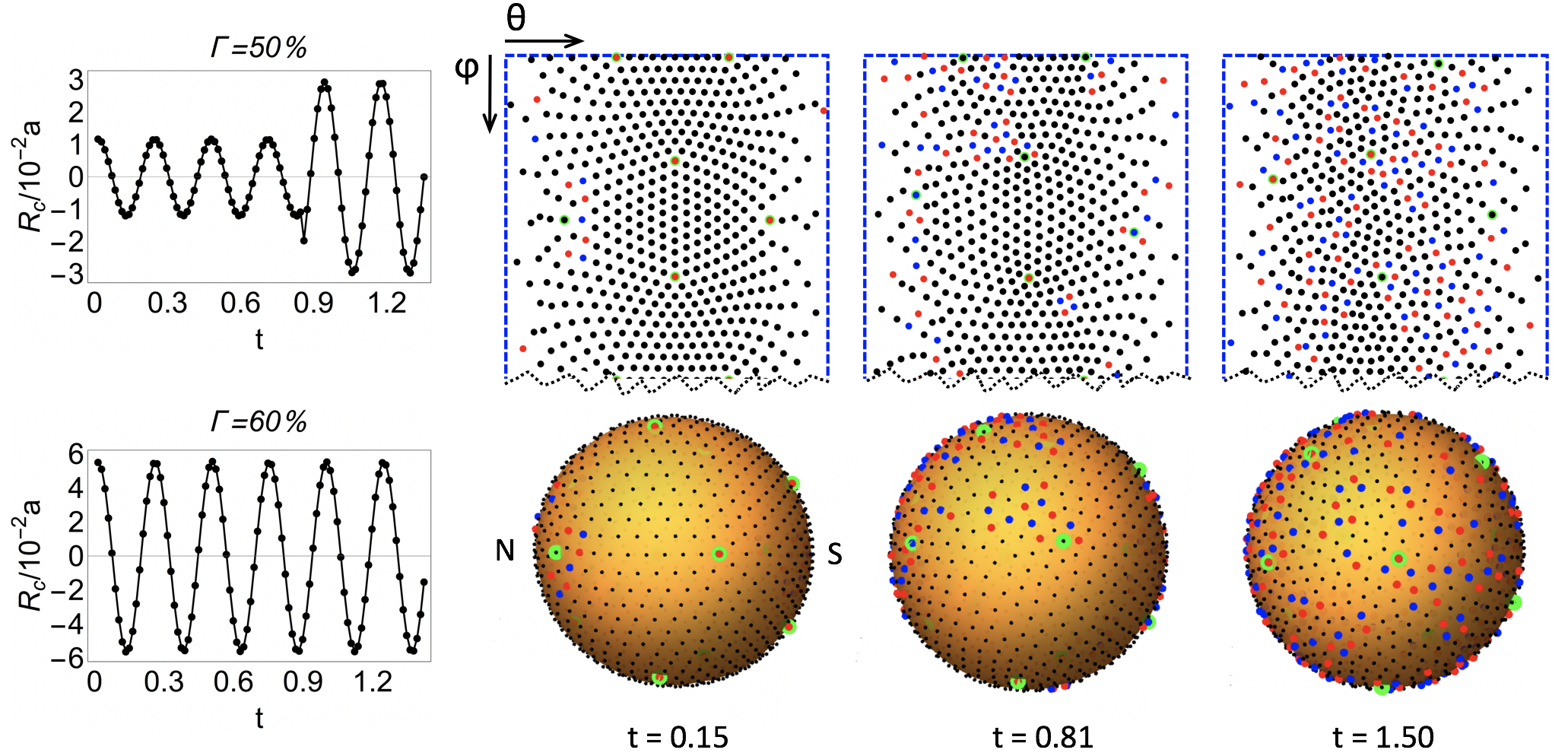}
  \caption{$\Gamma$-driven disruption of crystalline order in the spherical
  crystal. Left panels show the collective oscillation at $\Gamma=50\%$ and
  $\Gamma=60\%$. $R_c$ is the location of the center of mass on the z axis. The
  three snapshots of particle configurations at $\Gamma=60\%$ plotted in the
  spherical coordinates and on the sphere show the migration of the emergent
  defects (colored dots) from the poles of the sphere to the equatorial region
  during a few collective oscillations. Green dots represent the pre-existent
  twelve disclinations. Red and blue dots are five- and seven-fold
  disclinations. $(p, q)=(10, 0)$.  }
  \label{melting}
\end{figure*}

Spatiotemporal analysis of the velocity field reveals alternating inward and
outward movement of the five neighboring particles surrounding each
disclination. Such a breathing mode is crucial for fabricating the highly
symmetric velocity field. We quantitatively characterize the breathing mode by
introducing the quantity $r_B(t)$, which is the temporally varying average
distance from the disclination to the five neighboring particles. The
coincidence of the $r_B-t$ curves of the twelve disclinations in
Fig.~\ref{modes}(a) indicates the spontaneous synchronization of the local
breathing modes. And the transition to this ordered dynamical state is a fast
process (in comparison with the breathing period).  This salient feature is also
found in spherical crystals with distinct symmetries; a typical case of $q=2$ is
shown in Fig.\ref{modes}(b).

We further inquire under which conditions the breathing mode will emerge. We
examine spherical crystals of all kinds of symmetries for $q\in[0,p]$, and vary
the value of $\Gamma$ from $0.1\%$ up to $25\%$. It turns out that
both the frequency and amplitude (measured by the lattice spacing) of the
breathing mode are independent of the $\Gamma$ and $q$ values. It is noteworthy
that changing the $q$ value leads to the variation of the number of particles.
So the emergence of the breathing mode is also independent of
particle density, which is due to the lack of a length scale in the potential
$V(r)$.  Furthermore, we disturb the system in various ways, such as using a
series of random initial states with randomly distributed $\Gamma$, disturbing
randomly picked particles, disturbing only the disclinations, etc. In all these
cases, the system quickly enters the dynamical state of synchronized breathing
modes.

To explore the origin of the breathing mode, we perform normal mode analysis for
the elementary pentagonal configuration, and find that the radial eigenmode is
exactly the mode selected by the system to fabricate the breathing mode (see
Supplemental Material).
Furthermore, from the perspective of symmetry, such a mode preserves the local
fivefold rotational symmetry around the disclination.  As a numerical
observation to substantiate this viewpoint, we find that once an isolated
fivefold disclination is converted to a scar that is composed of alternating
five and sevenfold disclinations, i.e., the local fivefold symmetry is broken, the
breathing mode also vanishes~\cite{Bausch2003e}. Note that in this
disclination-scar conversion, the bond-orientational order is still
preserved~\cite{nelson2002defects}. From the perspective of energetics,
the symmetry-broken dynamical mode around the disclinations seems unfavored. We
design a symmetry-broken dynamical mode by moving a disclination and a randomly
picked undefected particle towards their neighboring particles respectively.  In
this process, the system energy of the former case increases much faster,
indicating that a symmetry-broken dynamical mode around the disclination is
energetically unfavored.

The double-peak structure in the red energy curves in Fig.~\ref{energy} implies
a new dynamical mode supported in the spherical crystal system. To identify this
mode, extensive data analysis of the numerical results leads us to examine the
trajectory of the center of mass of the system. It turns out that the
center-of-mass trajectory coincides with the two-fold axis of the
icosahedron spanned by the twelve disclinations that passes through the
midpoints of the two opposite edges. By
setting this line to be the z axis, we reveal a small-amplitude collective oscillation of
the system along the z axis, as shown in Fig.~\ref{modes}(c) and
\ref{modes}(d).  This dynamical mode is pervasive in spherical
crystals with distinct symmetries, with scars (which are line defects out of
isolated point disclinations), and even with disrupted crystalline order. In contrast, the
breathing mode is suppressed in the latter two cases.  The ubiquity and
the small-amplitude features of the collective oscillation mode imply that this
is a fundamental dynamical mode in the spherical crystal system. A systematic
survey in the parameter space of $q$ and $\Gamma$ (prior to disruption of
crystalline order) shows that the ratio of the frequencies of the
temporally varying kinetic energy, the breathing mode and the collective
oscillation mode is a constant:
\begin{eqnarray}
w_{E}:w_{B}:w_{O}=1:\frac{1}{2}:\frac{1}{4}.
\end{eqnarray}
In comparison with Fig.~\ref{energy}, we identify the collective
oscillation as the dynamical mode underlying the double-peak structure in the
energy curves.

We proceed to discuss the connection of the collective oscillation mode and
the disruption of the spherical crystal. The failure of the particles to be
restored at sufficiently large disturbance is the microscopic origin of the
$\Gamma$-driven disruption of crystalline order.  As a signal of phase
transition, we find the abrupt increase of the amplitude of the collective
oscillation, as shown in the left upper panel in Fig.~\ref{melting} when
$\Gamma=50\%$. The system is softened in the sense of the significantly
enhanced oscillation amplitude. What is the origin of the softening effect?
Frame-by-frame analysis of the Delaunay-triangulated particle configurations
shows that the softening of the system occurs exactly upon the appearance of
topological defects~\cite{nelson2002defects}.  In a static
system, the appearance of crystalline defects tends to reduce
stress~\cite{nelson2002defects, Audoly2010b}. Here, through the softening
phenomenon, we show the dynamic effect of topological defects.

From the characteristic snapshots in the disruption process of crystalline order
at larger $\Gamma$ as shown in Fig.~\ref{melting}, we see the migration of the
emergent defects (indicated by colored dots) from the poles of the sphere to the
equatorial region during a few collective oscillations. The connection of the
collective oscillation mode and the characteristic events in the disruption
process, such as the softening of the system and the global migration of
defects, may be attributed to its long-wavelength nature; long-wavelength
fluctuations make dominant contribution in phase
transition~\cite{nelson2002defects}. The snapshots in Fig.~\ref{melting} also
show that in the disruption process the relative positions of the preexistent
disclinations (in green dots) are subject to large deviation due to the
softening of the crystal lattice (see Supplemental Material for quantitative
analysis of the drift process of the disclinations).

In summary, by combination of numerical simulations and analytical normal mode
analysis, we have shown that topological defects in particle arrays can be a
crucial element in regulating collective dynamics and achieving dynamic order.
Specifically, we highlight two generic dynamical modes: synchronized breathing
modes around disclinations that induce highly symmetric singularity structures
in the velocity vector field, and collective oscillation with strong connection
to disruption of crystalline order. These results suggest an organizational
principle based on crystalline defects, and may inspire new strategies for
harnessing intriguing collective dynamics in extensive nonequilibrium systems.

This work was supported by NSFC Grants No. 16Z103010253, the SJTU startup fund
under Grant No. WF220441904, and the award of the Chinese Thousand Talents
Program for Distinguished Young Scholars under Grant No. 17Z127060032, and No.18Z127060018.

\end{document}